\documentclass[11pt,twoside]{article}
\usepackage{asp2004}
\usepackage{epsf}
\usepackage{psfig}
\usepackage{lscape}
\begin{document}

\title{Outflowing disk winds in B[e] Supergiants}


\author{Michel Cur\'{e}}
\affil{Departamento de F\'{\i}sica, Facultad de Ciencias, 
Universidad de Valpara\'{\i}so, Chile.}
\author{Diego F. Rial}
\affil{Departamento de Matem\'{a}ticas, Facultad de Ciencias Exactas y Naturales, 
Universidad de Buenos Aires, Argentina.}
\author{Lydia Cidale}
\affil{Facultad de Ciencias Astron\'{o}micas y Geof\'{i}sicas, Universidad Nacional de La Plata, Argentina.}

\abstract{
The effects of rapid rotation and bi--stability upon the density contrast between the equatorial and polar directions of a B[e] supergiant are investigated. Based on a new slow solution for different high rotational radiation--driven winds  and the fact that bi--stability allows a change in  the line--force parameters  ($\alpha$, $k$, and $\delta$), the equatorial densities  are about $10^2$--$10^3$ times higher than the polar ones. These values are in qualitative agreement with the observations. This calculation also permits to obtain the aperture angle of the disk.

} 
%
\section{Introduction}

Stellar winds play an important role in the evolution and the observed  physical properties of B[e] supergiants. Most of these early-type objects are IRAS sources. They show an optical spectrum dominated by metallic emision lines of permitted and forbidden transitions from singly ionized elements with FWHM $\sim$ 100 km s$^{-1}$. Their UV spectrum shows lines of superionized elements with large Doppler shifts.

The peculiar spectrum of B[e] supergiants has been interpreted by the presence of a slowly outflowing equatorial disk and a normal fast polar wind (Zickgraf et al.~1985). In order to explain the formation of this two-component radiation--driven  wind around early--type stars, Lamers \& Pauldrach (1991) have introduced the bi--stability mechanism.  Vink et al.~(1999) have shown that the bi--stability mechanism induced by rotation in line driven-winds is due to the radiative acceleration by iron, caused by the recombination of Fe IV to Fe III.

Pelupessy et al.~(2000) have calculated the density contrast (the ratio between equatorial and polar densities) in a B[e] supergiant for rotationally--induced bi-stability models applying multi--scattering line-force parameters above and below the critical temperature of the bi--stability jump ($\mathrm{T_eff} = 25,000\,$ K). The models were computed considering values of  
$\Omega = \mathrm{v_{rot}/v_{brkup}} \la 0.6 $, where $\mathrm{v_{rot}}$ is the equatorial rotational speed and $\mathrm{v_{brkup}}$ is the break--up speed, and the calculated ratio between equatorial and polar densities was about $\sim 10$, a factor $10$ times smaller than Bjorkman's (1998) calculations. 

However, B[e] supergiants represent a post-main sequence evolutionary phase of massive stars and are located near the Eddington 
limit (Zickgraf et al.~1986). Then, critical rotation speed is reached at a much lower stellar rotational speed. Velocities of about 200 km s$^{-1}$ would make the star rotate sufficiently close to the break-up speed to produce observable effects. Consequently, the wind characteristics near the equator are expected to differ from the polar wind.

Cur\'{e} (2004) proved that the standard solution (hereafter the fast solution) of the modified  CAK wind model (hereafter m--CAK, Friend \& Abbott 1986) vanishes for rotational speeds of $\sim$ 0.7 -- 0.8 $\mathrm{v_{brkup}}$, and there exists a new solution that is much denser and slower than the known standard  m--CAK solution. We will call it hereafter the {\textit{slow}} solution.\\

The aim of this work is to re-investigate the formation of an equatorial disk--wind for rapidly rotating B[e] supergiants, taking into account: 1) the fast and slow solutions of rotating 
radiation--driven winds that depend on the assumed rotational speed and 2) bi--stability line--force parameters.\\ 

In section \ref{resultados} we show the  density contrasts for a B[e] supergiant that result from 
m--CAK models with fast and slow solutions. The calculations of the aperture angle of the disk is 
presented in section \ref{aperture} The discussion and the conclusions are presented in sections 
\ref{diskusion} and \ref{conclu}, respectively.


\section{Wind model results \label{resultados}}

In order to investigate the influence of the rotation and the bi--stability jump on forming a disk--wind, we solve the non--linear momentum equation for the m--CAK wind in both polar and equatorial directions. Details and calculation methods used here are found in Cur\'e ~(2004). Since there is no calculation of the line--force parameters considering the 
slow solution found by Cur\'e ~(2004), we adopt the line--force parameters, $\alpha$ and $k$, published by  Pelupessy et al.~(2000) that 
were calculated for both sides of the bi--stability jump. We do not include the parameter
$\delta$ determined by Abbott (1982) since this parameter has very small effects on the 
density ratio $\rho_{e}/\rho_{p}$ (Cur\'e et al. 2005a). The line--force parameters are summarized in Table \ref{tabla1}.

We model a B[e] supergiant star with following parameters: $\mathrm{T}_{\mathrm{eff}}$ = 25,000 K, 
$\mathrm{M/M_{\sun}}$ = 30, $\log\mathrm{L/L_{\sun}} = 6$ 
and solar abundance. For the lower boundary condition for polar and equatorial directions, we set a constant value for the density in the photosphere, $\rho_{p}(R_{\ast}) = 5 \times 10^{-10}$ $\mathrm{gr \, cm^{-3}}$ (de Araujo \& de Freitas--Pacheco 1989).
%
   \begin{table}
      \caption[]{Bi--stability line force parameters}
         \label{tabla1}
     $$
         \begin{array}{p{0.1\linewidth}ccccc}
            \hline  \hline
            \noalign{\smallskip}
            T ${[\mathrm{K}]}$ & \alpha & & k & & \delta \\
            \noalign{\smallskip}
            \hline
            \noalign{\smallskip}
            30,000 & 0.65 & &0.06 & & \, 0   \\
	    17,500 & 0.45 & &0.57 & & \, 0   \\
            \noalign{\smallskip}
            \hline
         \end{array}
     $$ 
   \end{table}
   \begin{table}
      \caption[]{Parameters calculated for a star with $\mathrm{T_{eff}} = 25,000$ K, 
      $\log \mathrm{\,g}= 2.5$ applying  m--CAK models: Rotational parameter $\Omega$, terminal 
      velocity, $\mathrm{v_{\infty}}$ ($\mathrm{km\, s^{-1}}$) and $ r_{c}/R_{\ast}$, 
      the location of the critical point.}
         \label{tabla4}
     $$ 
	 \begin{array}{lcccccc}
            \hline \hline
            \noalign{\smallskip}
            & \Omega  & &v_{\infty} && r_{c}/R_{\ast}\\
            \noalign{\smallskip}
            \hline
            \noalign{\smallskip}
            pole    & 0.0&\,\,&679 && 1.14 \\
 	    \hline
            equator^{\mathrm{a}}  & 0.6 & &  253 &&  1.19 \\
            equator & 0.7 & &  178 && 4.19 \\
            equator & 0.8 & &  165 && 5.57 \\    
            equator & 0.9 & &  153 && 6.43 \\
            equator & 0.95 & & 147 && 6.78 \\            %
	    \noalign{\smallskip}
            \hline
         \end{array}
     $$ 
     \begin{list}{}{}
	\item[$^{\mathrm{a}}$] \small{Fast solution}
     \end{list}
   \end{table}

In this work we have not taken into account either the change in the shape of the star or 
gravity darkening (von Zeipel effect) or the modification of the finite--disk correction 
factor due to the rotation (Cranmer \& Owocki~1995 eq. [26], Pelupessy et al.~2000). However, we expect that these effects may have a small influence on the fast solution since we have not found large difference between our models (Cur\'e, et al.~2005a) and  Pelupessy et al.~(2000) results.

   \begin{figure}[!ht]
      \plottwo{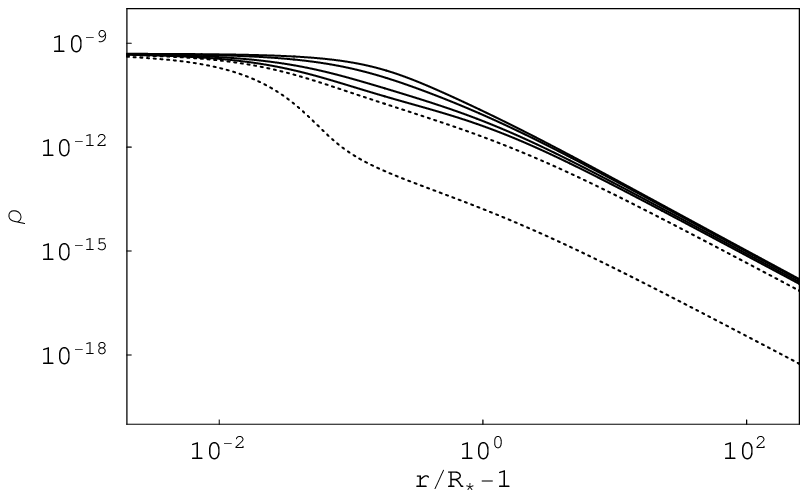}{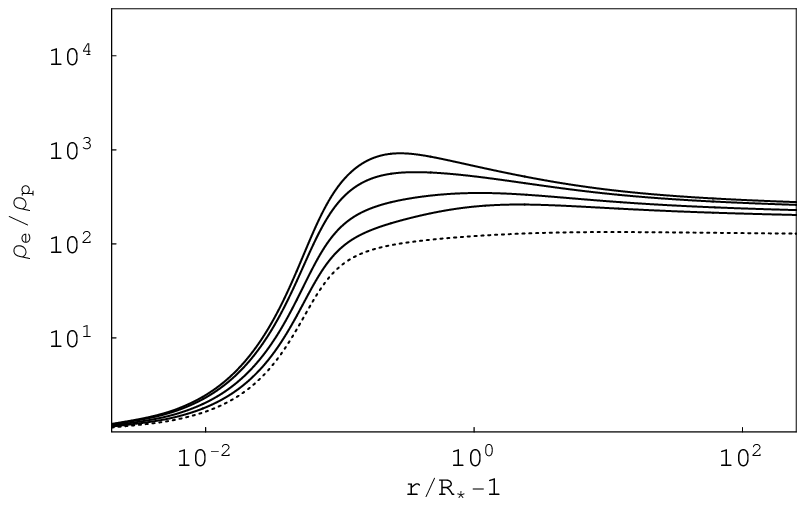}
      \caption{Left: density (in $\mathrm{gr\,cm^{-3}}$) versus $r/R_{\ast}-1$. Polar density 
               is in dotted--line; equatorial density for $\Omega=0.6$ (fast solution) is also in
               dotted--line and equatorial densities (slow solutions) for 
               $\Omega=0.7$, $0.8$, $0.9$, $0.95$ are in continuous--line, the higher is 
               $\Omega$, the higher is the density.
               Right: density contrast versus $r/R_{\ast}-1$, dotted--line is for $\Omega=0.6$ 
               (fast solution) and continuous--lines are for (slow sloutions) 
               $\Omega=0.7$, $0.8$, $0.9$, $0.95$. The higher  is $\Omega$, the higher is 
               the density contrast.
      }
      \label{fig3}
   \end{figure}

Figure \ref{fig3} illustrates density distributions and density contrasts for different 
rotational speed computed for m--CAK wind models and the set of line--force parameters given in Table \ref{tabla1}.

Due to the fact that the  existence of the fast or slow solution depends on the rotational speed, 
we also show in Figure \ref{fig3} the density profiles obtained; fast solution for $\Omega=0.6$ and  slow solutions for higher rotational speeds, $\Omega=0.7$, $0.8$, $0.9$, $0.95$.

The fast solution, which  is shown in Figure \ref{fig3} in dotted--line, yields to a density distribution that is lower than the densities attained with the slow solutions (continuous--lines) and higher than the polar density ($\Omega=0$). We obtain a density contrast of some hundreds for almost all the wind. The increase in the density contrast in the region close to the photosphere is due to the centrifugal force and the consequently higher mass--loss rate of the fast solution when rotation is included (Friend \& Abbott 1986).

Density contrasts reach values of around some hundreds to thousand for radii less than $\sim 2 R_{\ast}$ and an approximately value of some hundreds is maintained by the wind up to hundreds of stellar radii, almost independently of $\Omega$. This result concerning the disk behaviour is in qualitative agreement with the values estimated from observations by Zickgraf et al.~(1985,~1986,1992), Zickgraf~(1998), Oudmaijer et al.~(1998) and  Bjorkman~(1998).

\section{Aperture angle of the wind \label{aperture}}
In order to know the semi-aperture angle of the disk-wind, we start from the equator and calculate
up to which angle (from the equator) the slow solution exists, for greater values of the angle, only the fast solution exists. Therefore, we define the semi--aperture angle of the disk the {\it{last}} possible angle when the slow solution exists. Table \ref{tabla5} summarized the semi--apperture angle for the star used here for different assumed rotational speeds. The semi--aperture angle of the disk is larger as the rotational velocity increases.\\
\begin{table}
\caption[]{The semi-aperture angle of the disk}
   \label{tabla5}
$$
         \begin{array}{p{0.1\linewidth}cc}
            \hline  \hline
            \noalign{\smallskip}
            $\,\Omega $ & &\Theta_{\delta=0}\\
            \noalign{\smallskip}
            \hline
		0.7   & & 25\\ 
		0.75  & & 32\\
		0.8   & & 38\\
		0.85  & & 42\\
		0.9   & & 45\\
		0.95  & & 48\\
            \noalign{\smallskip}
            \noalign{\smallskip}
            \hline
         \end{array}
     $$ 
\end{table}

   \begin{figure}[!ht]
      \plottwo{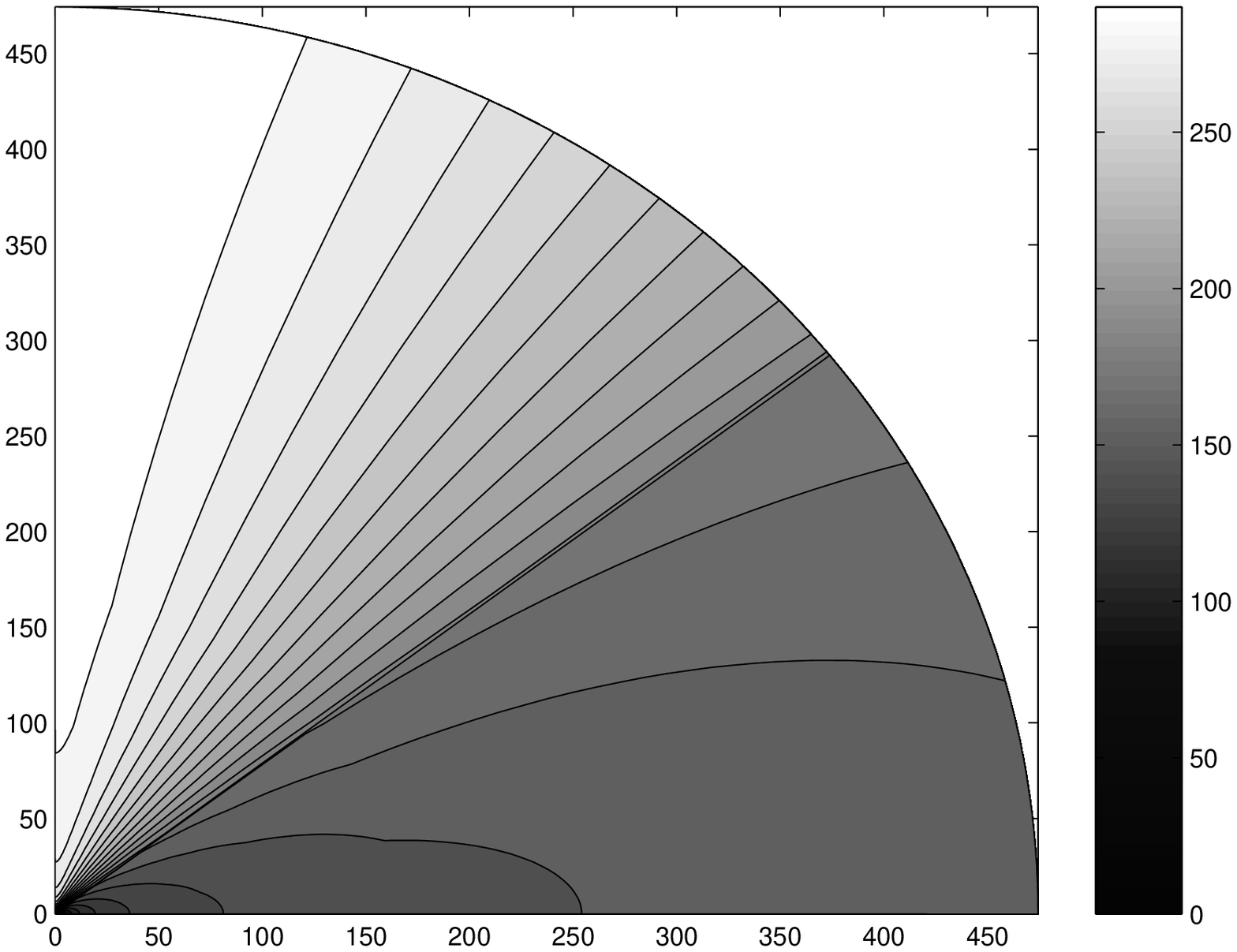}{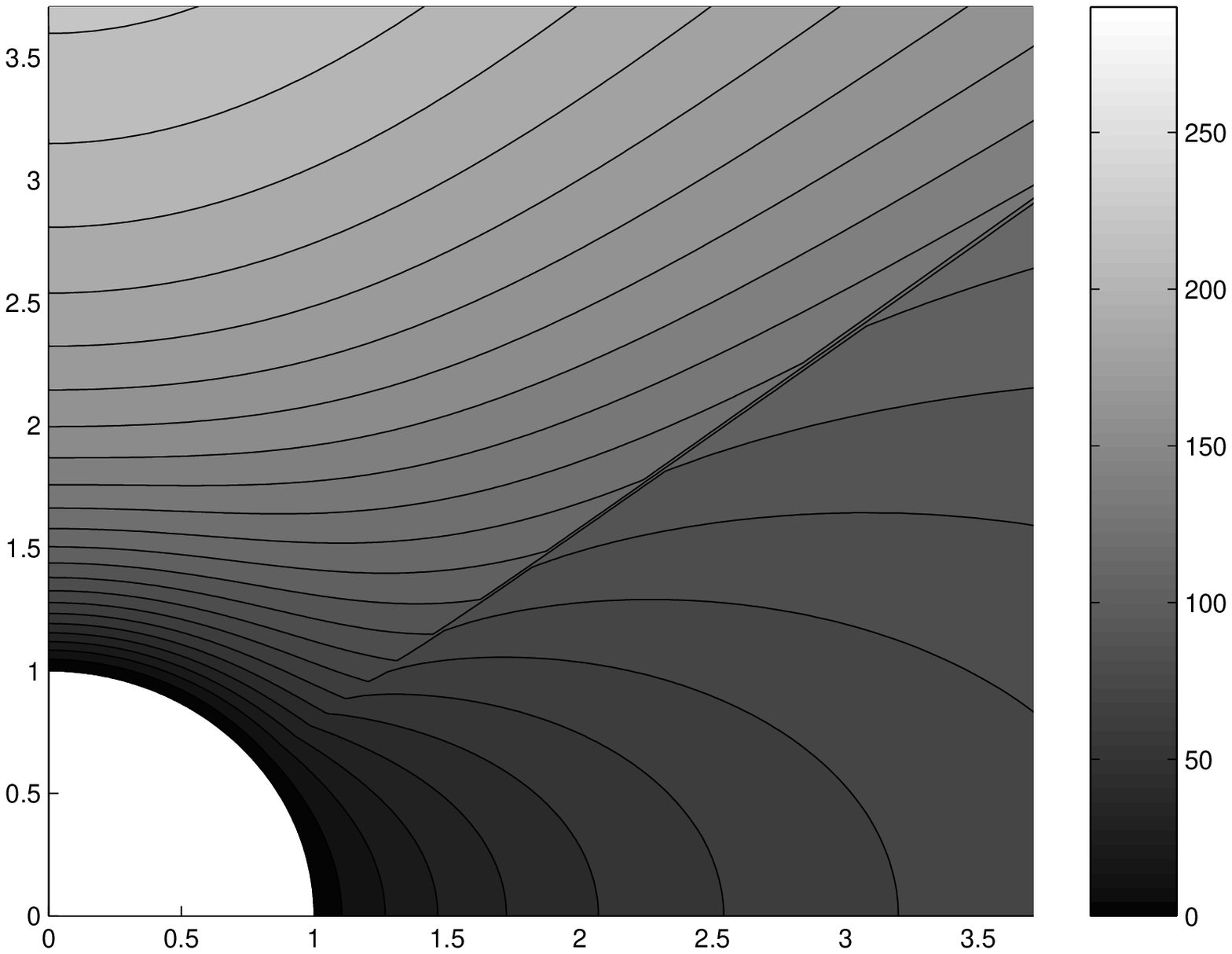}
      \caption{Left: velocity profile (in $\mathrm{km\,s^{-1}}$) versus $r/R_{\ast}$ for
               $\Omega=0.8$. The lines denote  iso--contour velocities.\,\,
               Right: Same as Left, but a zoom near the photosphere. 
      }
      \label{fig5}
   \end{figure}
%
\section{Discussion \label{diskusion}}

We want to stress the importance of the combined effect of slow and fast solutions
with bi--stable line--force parameters in forming an outflowing disk wind in B[e] 
supergiants. Density contrasts of the order of some $10^2$ up to large distances from the star are 
attained.
This theoretical value is in qualitative agreement with the values derived from the observations 
of the order of $10^2$--$10^3$ (Zickgraf et al.~1989, Zickgraf 1998 and references therein, 
Bjorkman 1998). The half-opening angle of the gaseous disk increases with the rotational velocity and its values are in the range 25 to 50 degrees. This angle depends strongly on the value of the star's gravity, the greater is the gravity ($\log\, \mathrm{g}$) the smaller is the disk angle (Cur\'{e} et al. 2005b).

Previous simulations of gaseous disk formation in rotating radiation--driven winds induced by 
bi--stability have underestimated the density contrast, mainly due to: a) The use of 
a $\beta$--field (with $\beta=1$, Lamers \& Pauldrach~1991) to describe the wind velocity 
profile, even for high rotational speeds where this approximation fails (Cur\'e~2004), and b) Pelupessy et al.~(2000) calculations, based in the fast solutions, were restricted to values of $\Omega \le 0.6$ and these rotation values are not high enough to develop a dense disk. 

A dense disk is formed when the slow solution starts to exist. For our test star, this occurs (numerically) for $\Omega \ge 0.7$. This condition is in qualitative agreement with the estimation of $0.74 \le \Omega \le 0.79$ by Zickgraf~(1998) in order to reproduce observable effects in the structure of stellar winds. However,  observational rotation speeds of B[e] supergiants have high uncertainties, because only a few stars show photospheric absorption lines appropiate for the measurements of $V\,\sin(i)$. The inferred observational value of $\Omega$ lies in the range $0.4$--$0.7$ (Zickgraf~1998).

Since most of the B[e] supergiants in the H--R Diagram are located below the bi-stability 
jump temperature ($25,000$K), in our conception, the theoretical explanation for the 
existence of a two--component wind model (Zickgraf et al.~1985) is due to the nature 
of the solutions of rapidly rotating radiation--driven wind. The change (jump) from the fast solution to the slow solution at some latitude yields to a two-component wind, where each solution structure has its own set of line--force parameters. This picture would be remarked for cases when the bi--stability jump is present. 

The semi--aperture angle of the disk is quite large when the star approachs to the limit of rotation velocity but this angle defines the gaseous structure. However, if a hot dust disk structure develops the dust disk would present a high concentration towards the equatorial plane.

Another important aspect to remark is the scarcity of self--consistent calculations of 
line--force parameters $k$, $\alpha$, $\delta$ for the m--CAK fast solution and the lack 
of calculations for our slow solution. The uncertainty in the values of the parameters is
reflected in the value of the terminal velocity, mass loss rate, as well as in the density 
contrast. Specifically, the predicted terminal velocities, see Table \ref{tabla4} 
are about two times greater than values inferred by observations (see Table [16] from Zickgraf 1998).

Therefore our results that combine fast and slow wind solutions are a first approximation to re-investigate disk formation in high rotating stars with radiation--driven winds. A detailed 
wind model needs a self--consistent line--force parameter calculations for both fast and slow solutions for a wind consisting of gas and dust components.

\section{Conclusions \label{conclu}}
We have revisited radiative driven wind models for a raplidly rotating B[e] supergiant 
($\Omega \ga 0.6$) assuming a change in the line--force parameters due to the 
bi--stability jump. The existence of slow and fast solutions in the model predicts 
density contrasts which are of the order of $10^2 - 10^3$ near the stellar surface 
(r $\la$  2 $R_{\ast}$), while outside they fall to values of about some $10^2$ 
and the disk extends up to $\sim 100$ stellar radii. We estimate a half-openning angle of the gaseous disk is about 40 degree for $\Omega=0.8$.

\begin{acknowledgements}
This work has been possible thanks to the research cooperation agreement UBA/UV, UNLP/UV 
and DIUV project 03/2005. 
\end{acknowledgements}

\end{document}